\documentclass[conference]{IEEEtran} 
\ifCLASSOPTIONcompsoc
  \usepackage[nocompress]{cite}
\else
  \usepackage{cite}
\fi
\ifCLASSINFOpdf
\else
\fi
\usepackage[colorlinks=true,
            linkcolor=red,
            urlcolor=blue,
            citecolor=blue]{hyperref}
\usepackage{amsmath}
\usepackage{mathtools}
\usepackage{relsize}
\usepackage{graphicx}
\usepackage[scientific-notation=true]{siunitx}
\usepackage{algorithm}
\usepackage{filecontents,lipsum}
\usepackage{cite}
\usepackage{comment}
\usepackage{multirow}
\usepackage{amsfonts}
\usepackage[noend]{algpseudocode}
\usepackage[utf8]{inputenc}
\usepackage{graphicx}
\usepackage{caption}
\usepackage{latexsym}
\usepackage{etoolbox}
\usepackage{subcaption, float}
\makeatletter
\usepackage[inline]{enumitem}
\def\BState{\State\hskip-\ALG@thistlm}
\makeatother

\AfterEndEnvironment{table}{\vskip-1ex}

\begin{document}

\title{Dialogue-Based Simulation For Cultural Awareness Training}

%
%

\markboth{IEEE Transactions on Medical Imaging}%
{Shell \MakeLowercase{\textit{et al.}}: Bare Demo of IEEEtran.cls for IEEE Journals}
%




\author{\IEEEauthorblockN{Sodiq Adewole\IEEEauthorrefmark{1},
Erfaneh Gharavi\IEEEauthorrefmark{1},
Benjamin Shpringer\IEEEauthorrefmark{2}, 
Martin Bolger\IEEEauthorrefmark{1},
Vaibhav Sharma\IEEEauthorrefmark{2},\\
Sung Ming Yang\IEEEauthorrefmark{2}, and
Donald E. Brown\IEEEauthorrefmark{1}\IEEEauthorrefmark{2}}

\IEEEauthorblockA{\IEEEauthorrefmark{1} Department of Systems and Information Engineering,
University of Virginia,
Charlottesville, VA, USA}

\IEEEauthorblockA{\IEEEauthorrefmark{3} School of Data Science, 
University of Virginia,
Charlottesville, VA, USA}

\{\href{mailto:soa2wg@virginia.edu}{soa2wg},
\href{mailto:eg8qe@virginia.edu}{eg8qe},
\href{mailto:bs2ux@virginia.edu}{bs2ux},
\href{mailto:meb2fv@virginia.edu}{meb2fv}, 
\href{mailto:vs3br@virginia.edu}{vs3br},
\href{mailto:sy8pa@virginia.edu}{sy8pa}, 
\href{mailto:brown@virginia.edu}{brown}\}@virginia.edu\vspace{-15pt}}

\maketitle

\begin{abstract}~Existing simulations designed for cultural and interpersonal skill training rely on pre-defined responses with a menu option selection interface. Using a multiple-choice interface and restricting trainees' responses may limit the trainees' ability to apply the lessons in real life situations. These systems, also rely on a simplistic evaluation model, where trainees' selected options are marked as either correct or incorrect. The model cannot capture sufficient information that could drive an adaptive feedback mechanism to improve trainees' cultural awareness. This paper describes the design of a dialogue-based simulation for cultural awareness training. The simulation, built around a disaster management scenario involving a joint coalition between the US and the Chinese armies. Trainees were able to engage in realistic dialogue with the Chinese agent. Their responses, at different points, get evaluated by different multi-label classification models. Based on training on our dataset, the models score the trainees' responses for their awareness of the Chinese culture. Trainees also get feedback that informs the cultural appropriateness of their responses. The result of this work showed the following; i) A feature-based evaluation model improves the design, modeling and computation of dialogue-based training simulation systems; ii) Output from current automatic speech recognition (ASR) systems gave comparable end results compared with the output from manual transcription; iii) A multi-label classification model trained as a cultural expert gave results which were comparable with scores assigned by human annotators.\vspace{5pt}
\end{abstract}

\begin{IEEEkeywords} 
Dialogue-based simulation, cultural-awareness training, intelligent tutoring system, ill-defined domain training, conversational agent, culturally-aware expert system, cross-cultural competence, adaptive feedback system.
\end{IEEEkeywords}

\section{Introduction}\label{sec:Introduction}
\vspace{-2pt}
Dialogue-based interactive simulation is a class of intelligent tutoring system (ITS) often designed for teaching awareness in domains that are not well structured. Examples include: interpersonal skills training \cite{lane2007intelligent}, cross-cultural competence \cite{lane2008getting, dillenbourg2002virtual, lane2013learning}, negotiation skills \cite{kim2009bilat, core2006teaching}, confrontation management \cite{kolkmeier2017moral}, foreign language learning \cite{swartz2012intelligent, wang2009computer, wik2009embodied}, and interviewing skills \cite{yu2019open}.

This paper describes the design and development of an intelligent tutoring system for cross-cultural training. Training to improve cross-cultural interactions is difficult but simulations and games designed for this task have proven to be more effective than traditional classroom instruction \cite{brown2019design}. However, most of these systems have used multiple-choice selection or menu-based interface to capture learners' actions \cite{georgila2019using, kim2009bilat}. Menu-based interfaces can be very restrictive and do not reflect the natural mode of human-to-human communication. Multiple choice interaction may also impact the educational efficacy of the training systems\cite{litman_itspoke:_2004, moreno_case_2001} by limiting the possibility of the learner being able to transfer any knowledge gained to a real world situation.

Also, the assessment models in  existing ITS use simplistic uni-dimensional categorization \cite{lane2008getting}. These models categorize learners' actions into one of few categories such as correct/incorrect or negative/neutral/positive. Such simplistic evaluations do not capture the multiple features and dimensions that characterizes interactions in a typical unstructured domains.

This paper expands upon existing ITS by incorporating dialogue-based interaction for cultural competency training. To build this new system required completion of the following tasks:
\begin{enumerate}
    \item Abstraction of cultural concepts into a computationally tractable feature space; 
    \item Training multiple expert models to evaluate and score users' responses in the simulation; 
    \item Integrating a speech recognition system as the input interface;
    \item Implementing an expert assessment model and an adaptive feedback mechanism that work together to provide feedback based on the trainee input; and 
    \item Designing and conducting an experiment to evaluate the usability and performance of the system.

\end{enumerate}

Since the goal of the our training simulation is to improve the cultural competency of soldiers working with their counterparts from the Chinese army, the ITS described here is built on a Disaster Management Exercise (DME) scenario involving a joint coalition of the US and Chinese army. 

\section{Related Work}\label{sec:RelatedWork}
There is a significant amount of literature related to the main elements of this project. By way of organization we have grouped this research into four major areas: Architecture; Interface; Data collection; and User response evaluation.

\subsection{Architecture}\label{subsec:architecture}
A common architecture of intelligent instructional systems consists  of four main components. The domain expert model; learner model; pedagogical model; and the user interface. The domain expert model contains the concepts, rules and problem solving strategies for the domain to be learned. It provides the standard for evaluating learner's responses. Expert knowledge can be represented in various ways, including network presentations \cite{gamboa_designing_2002}, behavior trees\cite{geng2011hybrid,chan1992curriculum}, and finite state machines\cite{yankovskaya1997finite}. Most instructional systems perform a range of actions such as providing instruction, hint and feedback to the learner \cite{vassileva1995reactive,gross_feedback_2012,gutierrez2011adaptive} during interaction. The expert model is used to analyze user's input in order to provide feedback on errors. The learner's model consist of dynamic representation of the evolving knowledge of the learner. This include aspects (variables) of the learner’s behavior and knowledge that have assumed effect on performance and learning. The Pedagogical model handles the planning and regulation of teaching activities. This includes making decisions on the sequence of activities and strategy to achieve the learning objectives. The plans of this module is based on observations received from the learner about his/her progress as defined in the domain expert model. The user interface is the connecting point between the tutor and the learner. It serves to collect the user’s input as well as render the actions of the tutor. It may come in different forms and media, including a web/mobile app interface or virtual-reality environments.

\subsection{Speech-based dialogue interface} \label{subsec:Interface}
Speech driven interface in human-computer tutoring systems have been shown to help improve learning\cite{litman_itspoke:_2004,moreno_case_2001} and also facilitates adaptation of instructional strategy based on perceived affect in learners' responses\cite{forbes-riley_designing_2011}. While human tutors can respond to the content of the learners' response and perceive their emotions\cite{litman_recognizing_2003}, existing menu-option selection interface in dialogue-based cultural training simulations limits this capability. The need to close this performance gap, as it applies to simulation systems to learn cultural awareness, coupled with the recent advances in natural language processing, automatic speech recognition (ASR) and text-to-speech systems, drove the investigations presented in this paper.

\subsection{Data collection}\label{subsec:review-data-collection}
Data-driven dialogue systems have mostly relied on Wizard-of-Oz (WOz) \cite{gratch2015negotiation} or role play simulation \cite{shepherd2010investigating} methods to gather data. The collected data is then used to train a natural language understanding model to recognize users' input to the system. Other works have approached data collection using crowd-sourcing through amazon M-turk\cite{sharma2019data}. Sharma V. et al. \cite{sharma2019data}, took a step further by applying a data augmentation technique to improve the sample size and class distribution of an originally crowd-sourced data.

\subsection{Response evaluation and feedback mechanism}\label{subsec:expert-models}
Evaluating the trainee's response, in an intelligent tutoring system, is the main task of the expert model. However, semantic representation and natural language understanding steps needs to be performed prior to this evaluation. Continuous bag of words, term-frequency inverse document frequency (TFIDF) or word vector representation are some of the techniques that have been proposed in the natural language processing literature~\cite{mateas2005structuring, joachims1996probabilistic, salton1988term, pennington2014glove}.

\vspace{-15pt}

Menu-based option selection systems map each pre-defined trainee's response to a corresponding performance measure and feedback \cite{sheridan2018investigating,georgila2019using,moenning2016developing}. These systems evaluate trainees' input using supervised machine learning classification algorithm that learns from training examples annotated by subject matter experts \cite{lane2008getting, lane2007intelligent, lane2013learning}. The trained models classifies new input as positive, neutral, or negative; Others use simpler binary category of positive or negative \cite{lane2008getting}. Applying this annotation method to the case of cultural training simulation means we will define a score as $1$ = Culturally inappropriate; $2$ = Neither appropriate nor inappropriate; and $3$ = Culturally appropriate.\\
This approach has a number of limitations for our system; i) the complexity of culture-sensitive interaction requires context to properly classify a dialogue. A single score is uninformative and also does not capture any context nor dialogue history; ii) Identifying the difference between the labels may also be a problem as an utterance could fall into more than one category\cite{lane2008getting}. This happens if a response contains tokens that qualifies it as an appropriate response but also contains an addendum that tends to change the meaning when the utterance is considered as a whole. Humans may still be able to separate such response, the model, however, will be uncertain as to which score to assign such utterance; iv) Adaptive feedbacks are constructed to help learners get better as they interact with the system. Any handcrafted feedback may be off point and not generalize to all possible expressions that the players could make. This may impact the user experience and not provide any information on where the trainee actually erred. This also makes constructing an adaptive feedback almost impossible; v) Lastly, a single score assignment model is not natural of the way human adjudge an utterance for cultural appropriateness. Even though there may be individual differences on interpretation, it is natural to decompose a sentence into components that fits different cultural features. In this paper, we extend this single annotation mechanism into a multi-label model where each label defines a feature that characterizes the utterance based on expectation of the target culture. This method facilitates the decomposition of possible user's responses into feature space that simplifies the natural language understanding task for the machine learning expert model. Similar to the information state update formulation in task oriented dialogue systems \cite{traum2003information,ko2015new,kang2010reliable,kim2018integrated}, this method also defines a structure where cultural training dialogues becomes easier to implement in addition to facilitating adaptive feedback. Adapting instruction, hint and feedback\cite{vassileva1995reactive} to a trainee's performance in a simulation environment may lead to improved awareness of the target domain \cite{georgila2019using,lutticke2004problem} and also allows them to know how to subsequently provide more appropriate response \cite{gutierrez2011adaptive}\newline

\section{Methodology}\label{sec:Method} 
\subsection{Data collection, annotation and scoring methods}\label{subsec:data}
In an earlier work \cite{sharma2019data} we described the foundation for our data collection effort. Based on the dialogue designed by the Chinese culture experts\cite{sheridan2018investigating}, players' in the simulation are evaluated at fourteen (14) different points during the interaction. In order to train the expert models for this task, we crowd-sourced data using Amazon mechanical turk (m-turk). Three approaches were used to prompt m-turk users; i) using context from the scenario, they were asked to provide their own responses to the avatar comments; ii) using feedback provided by the Chinese culture experts in our multiple-choice version\cite{sheridan2018investigating}, they were asked to construct alternative responses; and iii) rephrasing the multiple-choice responses in the earlier version \cite{sheridan2018investigating} of the simulation. The collected data was  annotated by a minimum of two (2) annotators. The annotators had an average inter-rater reliability of 66\%\cite{sharma2019data}. The next section will detail how the annotated data was used in training the models to evaluate and score trainees' responses in the simulation. The scores assigned by the model are then mapped to the abstracted concepts. This mapping is used to construct a feedback that informed the players of their errors.\\

\subsubsection{Cultural Concept abstraction}
In \cite{sheridan2018investigating} subject matter experts provided feedback for each pre-defined option in the system. Using this feedback we developed binary features by abstracting the concepts that the Chinese culture experts used to determine the appropriateness of a response. These features allowed us to cast the problem as a multi-label machine learning classification problem.\\

\textit{Multi-label annotation} assigns a score vector with each component being mutually exclusive as follows;
\begin{equation}
    score = [x_{i}, . . ., x_{k}] : x_{i} \in \{0,1\}
    \label{score_vector}
\end{equation}

This approach makes more sense for a cultural training simulation because; i) In human-to-human interaction, an inappropriate utterance may not be completely inappropriate as there might be elements of such utterance that may still match the interlocutor's expectation. By decomposing the interlocutor's expectation into a set of features, the model will be able to separate an utterance based on the natural language representation as discussed in section \ref{subsec:expert-models}, and decide whether it is a hit or a miss. ii) Since context is also very important, the features can be defined to implicitly capture what makes an appropriate response within a particular context.

\vspace{5pt}

Table \ref{table:1} shows example response in one of the sections in the simulation where the player tries to greet the Chinese officer. The features in this section are as follows; \newline
\textbf{ A:} Greets officer by saying: "Hi", "Hello" or "Good morning"; \newline
\textbf{ B:} Avoids asking about officer's welfare on first meeting; and \newline
\textbf{ C:} Uses an honorific expression \newline

\begin{table}[!htb]
    \centering
    \begin{tabular}{p{3.5cm}|p{1.0cm}|p{1.0cm}|p{1.0cm}}
    \multicolumn{4}{c}{\textbf{Example of Cultural-feature abstraction}} \\
    \hline
    \textbf{Example response} & \centerline{\textbf{A}} & \centerline{\textbf{B}} & \centerline{\textbf{C}} \\ \hline
    Good morning captain Wang, how are you doing? & \centerline{1} & \centerline{0} & \centerline{0}\\ \hline
    Good morning captain Wang, how are you doing today? It's an honor to be here.  & \centerline{1} & \centerline{0} & \centerline{1}\\ \hline
    Hello captain Wang, it's an honor to meet you today. & \centerline{1} & \centerline{1} & \centerline{1}\\ \hline
    \end{tabular}
    \vspace{0.05cm}
    \caption{Cultural concept abstraction}
    \label{table:1}
\end{table}

This feature-based utterance label allows us to automate feedback construction such that each response is mapped to the features and a feedback could be constructed based on score received from the expert model.

\subsection{System Architecture}
\begin{figure}[ht]
 \centering
 \includegraphics[width=0.45\textwidth]{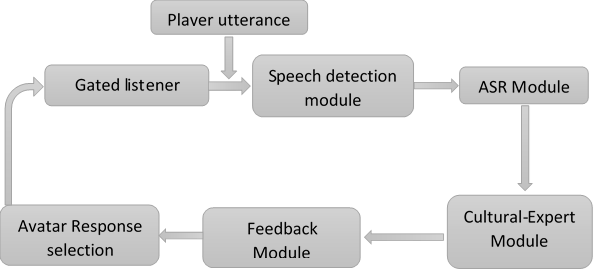}
 \caption{Simulation system architecture}
 \label{fig:architecture}
\end{figure}

Figure \ref{fig:architecture} shows the architecture of the simulation. The modular architecture allowed for easy integration, troubleshooting and error correction. We will describe the interface and each of the components in more detail below;

\subsubsection{User Interface}\label{subsec:interface}
\begin{figure}[ht]
 \centering
 \includegraphics[width=0.45\textwidth]{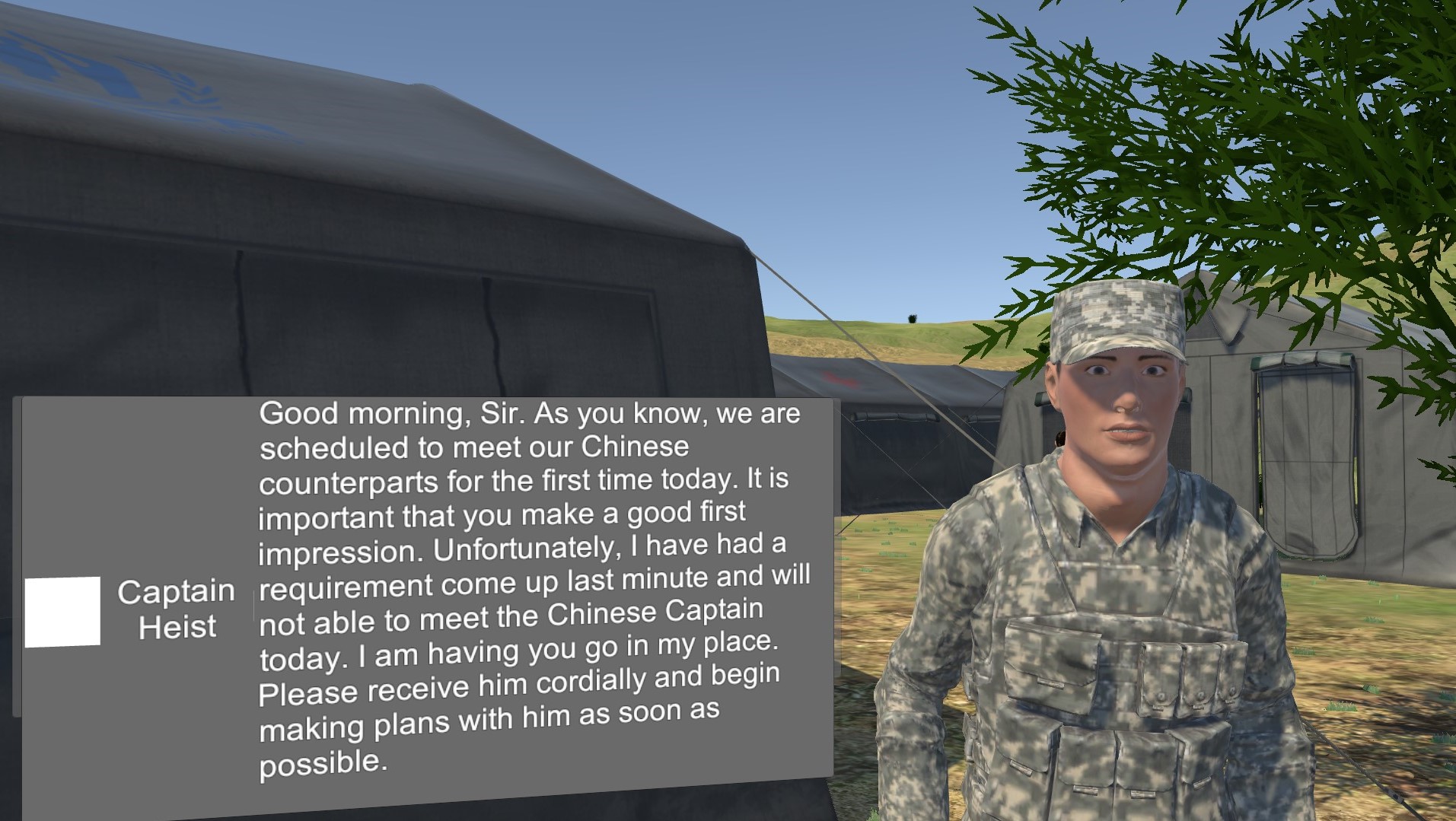}
 \caption{virtual simulation environment}
 \label{fig:simulation}
\end{figure}

\textit{Virtual environment:} The simulation environment was designed using the Unity game engine. Characters were modeled and rigged using Adobe fuse while animation was done using Mixamo 3D character animator. Lip-syncing and facial animation helped bring the characters to life and created a more realistic interaction in the virtual environment. Figure \ref{fig:simulation} shows one of the American soldier character in the virtual environment.\\

\textit{Voice Detection (VDM) and Automatic Speech Recognition (ASR) Modules:} The voice detection module is a gated listening mechanism that manages turn taking between the player and the avatar. It uses a sampling technique \cite{ramirez2007voice} to detect when the player starts to speak so as to begin the recording. Once the player completes their response, the recorded audio is passed to the ASR system for speech to text transcription. When the avatar is speaking, the listening gate is shot until the feedback is completely provided to the player. We found better transcription accuracy, using this approach as compared with direct streaming to the ASR system. Taking recording of the audio by section also allowed to evaluate effect of the error of the ASR on the performance of the model in properly scoring the player's input. We will describe this in more detail. We used the Google speech-to-text API for the speech transcription.\\

\subsubsection{Cultural-Expert Module} 
\begin{figure}[ht]
 \centering
 \includegraphics[width=0.45\textwidth]{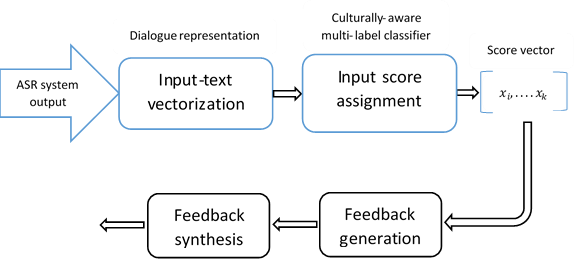}
 \caption{Cultural-expert module}
 \label{fig:expert_model}
\end{figure}

This module houses the pre-trained input-vectorizer and the multi-label classifier models as shown in figure \ref{fig:expert_model}. We trained multiple multi-label classifier models as our domain expert to evaluate and score the player's response for each section. We considered the problem of recognizing whether a user's utterance is culturally appropriate or not as a multi-label text classification problem\cite{mccallum1999multi,read2011classifier} as described in section \ref{subsec:expert-models}. Each user utterance is scored based on the defined features for the section. Since the goal of the simulation is to improve cultural awareness of soldiers working with their counterparts from the Chinese army, we built the scenario around a Disaster Management Exercise (DME) involving a joint coalition effort between the US and the Chinese army. Chinese culture experts designed a dialogue tree structure with fourteen (14) strategic points where the players' responses are evaluated for cultural awareness. For improved accuracy, each of these evaluation tasks are handled by separate evaluation models and trained on separate dataset as described in section \ref{subsec:review-data-collection}.

\textit{Input representation}
After experimenting with different dimensions of probabilistic semantic representation such as pretrained word vectors \cite{pennington2014glove},  bag of words (BOW) and Latent Dirichlet Allocation (LDA) \cite{blei2003latent}. The term frequency-inverse document frequency (TFIDF) representation gave best performance\cite{salton1988term} iin our model. Each evaluation points also had a pre-trained vectorizer that outputs the TF-IDF vector representation for each data point in our dataset. We used the same vectorizers in the simulation to take the text output from the ASR system to produce the corresponding vector representation. The architecture for this module is as shown in figure \ref{fig:expert_model}\\

It is important that the expert module has the correct text to score and so it is important that the output of the ASR system is as accurate as possible. We, therefore, set a confidence threshold for the output from the ASR system such that if the confidence is below this $\alpha$ threshold, the avatar will ask the player to repeat him/herself. 

The ASR module is configured to output both the transcribed text and the confidence parameter $\alpha$ for the recognized speech output. First, we tried different values of $\alpha$ to balance the trade-off between frustrating users and the accuracy of the expert model. This allows the speech recognizer to only pass output above $\alpha$ to the expert module. If the ASR module confidence level is below this threshold, the dialogue loops back and the avatar would ask the player to repeat what (s)he has said again.\\

\textit{Classifier models:} The models were trained on our annotated dataset of data as detailed in \cite{sharma2019data}. We experimented with different classifier models as well as different combinations of the input vector dimensions to determine the best performing model. We evaluated the performance of the models based on the F1-score, precision and recall. We obtained the best performance with k-nearest neighbours (KNN), RandomForest and multi-layer perceptron (MLP) classifier models. RandomForest and multi-layer perceptron (MLP) models were integrated into the final simulation because, while KNN had a good performance, the computational cost during the simulation was too expensive and the time taken for the player's response to be evaluated makes the deployment less suitable for a real-time interactive simulation.
Our evaluation metric was carefully chosen for reasons that captures both pedagogical objective as well as the user-experience considerations. Below, we detail suitability of each of these metric;

\textit{Precision:} is calculated as the ratio of the true positive (TP) to the sum of the true positive (TP) and false positive (FP). High precision indicates high TP and low FP which means that the expert-module mostly recognizes when the users make the appropriate. Also, low FP means the players were not wrongly assigned a score when they gave an inappropriate response. This is bad for pedagogical efficacy of the system since the users will not be corrected for what they did wrongly. Minimizing the FP is important for the pedagogical efficacy of our system.\\
\textit{Recall:} is the ratio of the TP to sum of TP and false negatives (FN). High recall indicates high TP to FN ratio which is desirable for a good user experience in the simulation. This means that the classifier mostly recognizes when the users gives the appropriate response and scores them properly for it. Since the feedback received is based on what the output score is, low false negative means the users are not frustrated by the system correcting them even when they make an appropriate response. We wanted a good recall rate so players will not receive feedback that says their response is not appropriate, even when it is. Inaccurate feedback is bad for educational effectiveness of the interaction as players may get confused as to what actually defines an appropriate response.\\
\textit{F1 score:} is the harmonic mean of precision and recall metrics. The F1 score is an important metric in this system as it aggravates the lower of the outlying value of the precision and recall metrics. It offers information as to the trade-off we are making between pedagogical efficacy and the user experience of the simulation. Precision and recall can individually be maximized by minimizing false positives and false negatives respectively.\\
\textit{ASR Word Error Rate:} measures the performance of the speech recognition module. This metric will help us understand and compare the performance of the expert models. Since the model performance is dependent on whether or not it got the correct token from the ASR system. If the expert module is not finding the token that identifies a user response as appropriate due to error from the speech recognizer, the expert module will consider this response as inappropriate and may assign a score of zero for that feature.\\

\subsubsection{Pedagogical module:} Two pedagogical components were implemented to help players navigate interactions with the avatars. A simulation guide help players navigate the interaction by providing background information about the non-player characters. The feedback module informs the players of what was right and wrong about their response. We will describe both components in detail below;

\begin{figure}[ht]
    \centering
    \includegraphics[width=0.45\textwidth]{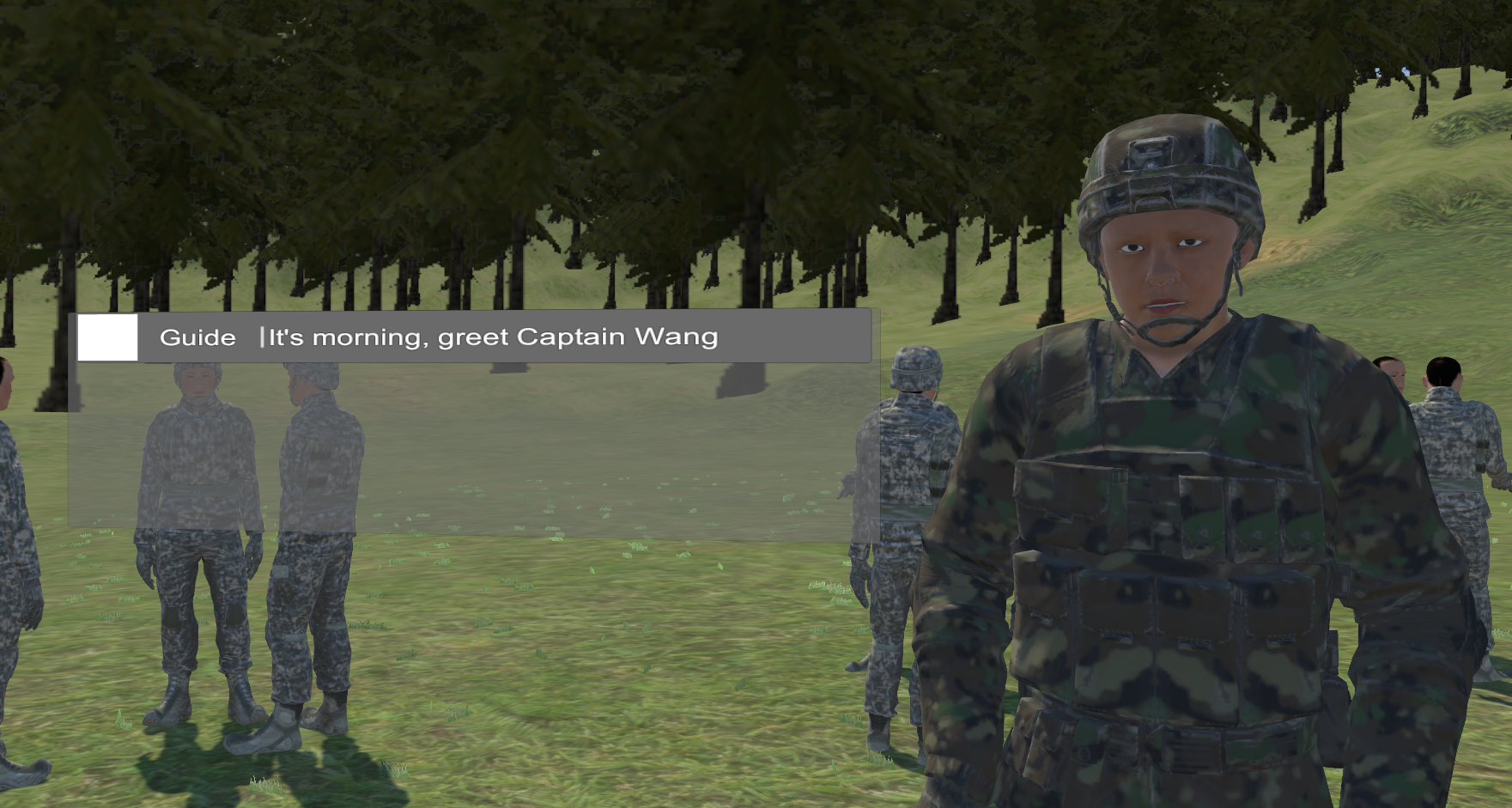}
    \caption{Simulation guide}
    \label{fig:CaptainWang}
\end{figure}

\textit{Simulation guide:}
The simulation guide is a pre-recorded audio with corresponding text display as shown in both figures \ref{fig:simulation} and \ref{fig:CaptainWang}. This provides the players with context of the scenarios and help guide them by providing additional information that the players need to properly construct their responses. This is important for several reasons; i) it helps them clarify and contextualize what was said by the avatar and also connects it to the history of the conversation. This will help the players recall the history of the dialogue in the previous scenes and connect to the current one;  ii) it helps narrow the space of potential response so the players could articulate their response within a shorter time; iii) it also helps the flow of the interaction when the player are able to respond promptly; iv) It increases the players immersion in the scenes and also their likelihood of making the proper response; and v) For non-military players who may not be familiar with the scenario, the guide helps them understand the context of what they needed to deal with.

\textit{Adaptive feedback generation} Adaptive feedbacks based on intelligent error analyses of learners’ solutions can help achieve correct response with minimal error as learners progress through the simulation \cite{gutierrez2011adaptive,lutticke2004problem}. This is especially useful for our simulation because, it help players continuously improve their responses as they try to implement the corrections in subsequent turns\cite{kolb2014experiential}. Our adaptive feedback mechanism uses the output score-vector as shown in equation \ref{score_vector} from the expert module and maps it to the domain feature space of the dialogue. Depending on the score output, the feedback generator constructs a feedback that informs the player of what an appropriate response should be and informs them of what they did right and tries to correct what they could improve upon in subsequent dialogue.

An example feedback for the first response from table \ref{table:1} which is an example utterance that a player could say when meeting the Chinese captain for the first time; (i.e.“Good morning captain Wang, how are you doing?”). The domain-expert module returns a score vector of [1,0,0] based on the following features;\\
\begin{flushleft}
(i)  “Greets the officer”  - [1] \\
(ii) “Avoids asking about officer's welfare” - [0]\\
(iii) “Uses an honorific expression” - [0] \\
\end{flushleft}

\begin{flushleft}
the feedback generated is as follows:\\
\end{flushleft}

\textit{“a culturally appropriate response in this section should include greeting the officer, avoiding asking about the officer's welfare on a first meeting, and using an honorific expression. From your response, you succeeded in greeting the officer, but your response could be improved by avoiding asking about the officer's welfare on a first meeting and using an honorific expression.”}\\

The generated feedback text is passed to the feedback speech synthesizer to get converted to audio format and played to the players before proceeding to the next turn.

\section{Experiment}
We recruited 18 subjects to test the usability and performance of the simulation. We evaluated the performance of the models and the overall system as a potential training tool for cultural awareness. Participants were mostly undergraduates from the school of engineering at the University of Virginia. The objective of our simulation system design was to explore culturally-aware data-driven model that evaluates the speech-based dialogue between the user and the computer avatar. Our experiment did not cover testing learning improvement in the target culture as this was outside the scope of this paper. The experiment was approved by the IRB and all participants provided their consent for the data that was collected.

Our hypotheses are as follows: \newline
\textbf{\textit{H1:}} The expert model will not assign scores comparable with human adjudged scores based on the players' responses in the simulation.\newline
\textbf{\textit{H2:}} The word error rate of the speech recognizer does not have any impact on the performance of the expert model when compared with manually transcribe input.\newline
\textbf{\textit{H3:}} The adaptive feedback mechanism aids the user in making more culturally appropriate responses as they progressed through the interaction. \newline
\textbf{\textit{H4:}} An integrated interactive simulation system with speech-driven input designed for cultural awareness training created room for more naturalistic interaction with better user experience.

To analyse the performance of the cultural-expert models, we collected the text transcript of the participants' responses as well as the audio recordings. This logged text also include the score assigned by the expert module and the ASR confidence level.
For the user experience, we administered questionnaires to the participants to understand what went right and how the system can be improved.

\section{Results}
The first five subjects’ data were removed because the system did not use the final versions of our classification models. This leaves the data for 13 participants. The system froze twice and the data collected for the two (2) participants was omitted in the analysis as well, leaving data for eleven (11) participants.  

\subsection{Expert Model performance}
The recorded audio responses were manually transcribed, the result was then annotated by two human annotators. The annotators had inter-rater agreement of 78\% and the score assigned was compared with the output scores from the expert models. Using the score from the human annotators as a baseline, we show the performance of the expert models in table \ref{tab:model_performance} below. The model-assigned scores was compared with the base-line to calculate the F1-score, recall, and precision. We also considered the ASR system word error rate to explain in part what was observed in the performance of the models.

\begin{table}[ht]
    \centering
    \begin{tabular}{|p{1.0cm}|p{1.0cm}|p{1.0cm}|p{1.0cm}|p{1.0cm}|p{1.0cm}|}
    \multicolumn{6}{c}{\textbf{Performance of cultural-expert module}} \\
\hline
\textbf{Model Id} & \textbf{No of features} & \textbf{F1 (\%)} & \textbf{Precision (\%)} & \textbf{Recall (\%)} & \textbf{ASR WER (\%)}\\ \hline
        1 & 3 & 94.1 & 88.9 & 100.0 & 63.2 \\ \hline
        2 & 3 & 80.0 & 69.2 & 94.7 & 25.4 \\ \hline
        3 & 2 & 93.0 & 87.0 & 100.0 & 17.3 \\ \hline
        4 & 3 & 75.0 & 68.2 & 83.3 & 19.7 \\ \hline
        5 & 2 & 79.1 & 77.3 & 81.0 & 15.0 \\ \hline
        6 & 3 & 86.7 & 96.3 & 78.8 & 17.3 \\ \hline
        7 & 3 & 63.4 & 65.0 & 61.9 & 14.6 \\ \hline
        8 & 4 & 52.9 & 45.0 & 64.3 & 15.3 \\ \hline
        9 & 3 & 80.0 & 72.0 & 90.0 & 8.6 \\ \hline
        10 & 2 & 88.0 & 100.0 & 78.6 & 19.3 \\ \hline
        11 & 1 & 88.9 & 80.0 & 100.0 & 4.4 \\ \hline
        12 & 2 & 83.3 & 100.0 & 71.4 & 19.9 \\ \hline
        13 & 3 & 89.4 & 95.5 & 84.0 & 19.2 \\ \hline
        14 & 2 & 76.9 & 66.7 & 90.9 & 16.4 \\ \hline
        \multicolumn{2}{|c|}{\textbf{Mean}} & \textbf{80.7} & \textbf{79.4} & \textbf{84.4} & \textbf{19.6} \\ \hline
        \multicolumn{2}{|c|}{\textbf{Std. Deviation}} & \textbf{11.4} & \textbf{16.1} & \textbf{12.6} & \textbf{13.5} \\ \hline
    \end{tabular}
    \vspace{0.1cm}
    \caption{Performance of cultural expert module}
    \label{tab:model_performance}
\end{table}

\begin{figure}[ht]
    \centering
    \includegraphics[width=8.5cm, height=6cm]{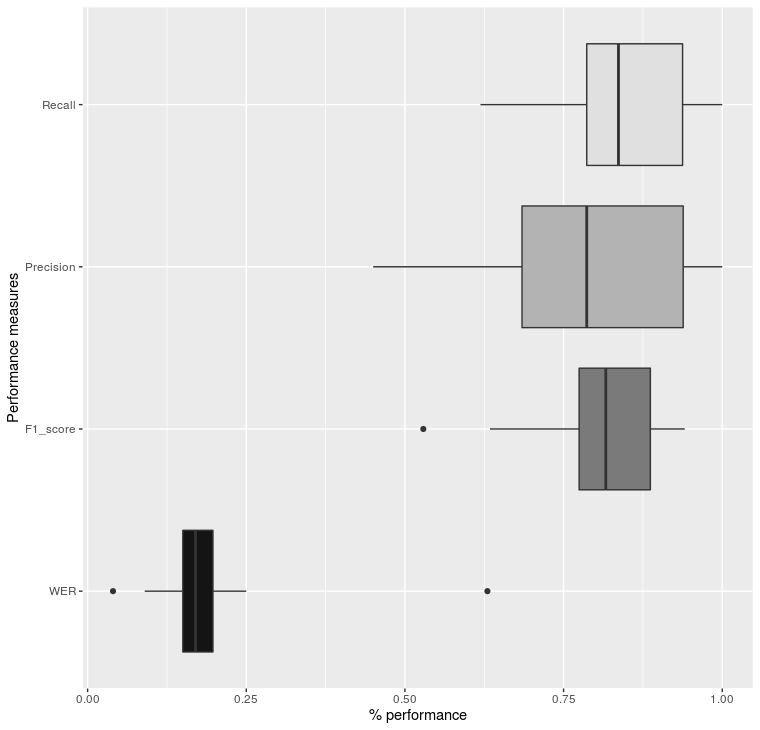}
    \caption{Model performance and ASR word error rate}
    \label{fig:model_performance}
\end{figure}

\vspace{5pt}

Figure \ref{fig:model_performance} shows the plot of the model performance in table \ref{tab:model_performance}. The expert models had average F1-score of 80.7\%, average precision of 79.4\% and recall of 84.4\%. The average WER of the ASR system is 19.6\%. 

\vspace{5pt}

\textit{\textbf{H1:}} The F1-measure of the expert models as shown in table \ref{tab:model_performance} shows average performance of 80\%. Except for model \textit{7} and \textit{8} with \textit{3} and \textit{4} features respectively, other models had F1-score above 75\%. Models \textit{7} and \textit{8} had more binary features resulting in ($2^{k}$) classes into which every response could fall into. From this we see that the number of features impacts the performance of the models as the classification task gets more difficult with increasing number of features. \newline

\vspace{5pt}

The high WER for section 1 is because participants were mostly mentioning the name of the Chinese avatar ("Captain Wang") with whom they were speaking and the ASR system struggled to properly recognize it. Training the ASR system on our own dataset may have improved the performance in this section. With average precision of 79.4\% the ratio of true positives (TP) to false positives is well above average which shows that the model does not assign score when it receives an inappropriate response from the players. Similarly, average recall of 84.4\% showed that the models TP to false negatives (FN) is around 0.84. This also indicates positive user experience performance as the players do not get corrected for what should pass as a culturally appropriate response. Both measures shows high TP values which is good for the pedagogical efficacy of the simulation. Since low recall values indicates higher FN, models 7 and 8 with the least recall means that players were, on average, corrected more than a human evaluator would have corrected similar utterance. Based on F1-score (t(13) = 26.49, p=1.07e-12), we have enough evidence to conclude that the performance of the expert module was comparable with that of human expert using manually transcribed data.

\textit{\textbf{H2:}} We performed simple regression analysis to analyse the relationship between the WER the F1-score. Our result [t(13) = 1.146, p = 0.274] showed no significant (p>0.05) relationship exist between the word error rate of the speech recognizer and the performance of the expert models. Similarly, the performance of the ASR system had no impact on the other metrics (Precision and Recall) of the expert models. This shows that, the tokens needed to score the participants' responses were mostly properly recognized. For example, model-1 where participants get score when they greeted the army officer that they were interacting, tokens such as "Good morning", "hi", "hello" were mostly correctly recognized and appropriately scored by the model. However, a multiple regression analysis using the number of features together with the WER to predict the F1-score of the models showed that both variables have joint significant effect [F(2,11) = 6.179, p=0.016] on the performance of the models. Therefore, as the number of features increases, the effect of the speech recognizer error tend to begins to have effect the model performance.

\textit{\textbf{H3:}}
Using \textit{average aggregate scores across sections} (equation (\ref{proportion of aggregate average score})), we evaluated the performance of the participants as they progress through the simulation. We hypothesized that providing an adaptive feedback to the players based on the model assigned score will improve their performance as they progress through the interaction. Natural human-to-human interaction does not usually provide explicit feedback to help players identify errors in their responses. However, it was still necessary to provide feedback to participant to help them learn of their mistakes based on the norm in the target culture.

\begin{equation}
    Score_{avg} = \frac{1}{N}\sum_{j}^{N}{(\frac{1}{k}\sum_{i}^{k}{x_{i}})}
    \label{proportion of aggregate average score}
\end{equation}

From equation (\ref{proportion of aggregate average score}), $N$ is the number of participants and $k$ is the number of cultural concept that each model was evaluating in the users' utterance; $k$ is defined for each model as the number of labels/features as indicated in table \ref{tab:model_performance}.

\begin{figure}[ht!]
    \centering
    \includegraphics[width=0.45\textwidth]{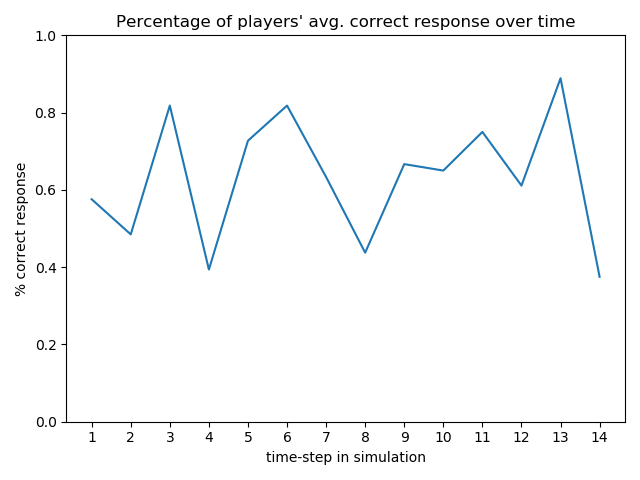}
    \caption{Participants average score}
    \label{fig:average_score}
\end{figure}

From Figure \ref{fig:average_score} above, we did not observe any performance improvement in the aggregate score received by the participants during the interaction. This is may be due to the independence of the features at each evaluation points. Also, most of the participants have not had any prior experience using such simulation system for cultural training and this may impact their performance as a first timer. The participants may also find it difficult remembering what corrections they received in previously sections. Also, lessons learnt in one simulation run can be easily utilized experientially in a subsequent one but not necessarily during the same interaction. Players would have had time to internalize the corrections from the first experience and if they have to go through the simulation on a second run, their experience may be better.

\begin{figure}[ht]
    \centering
    \includegraphics[width=0.45\textwidth]{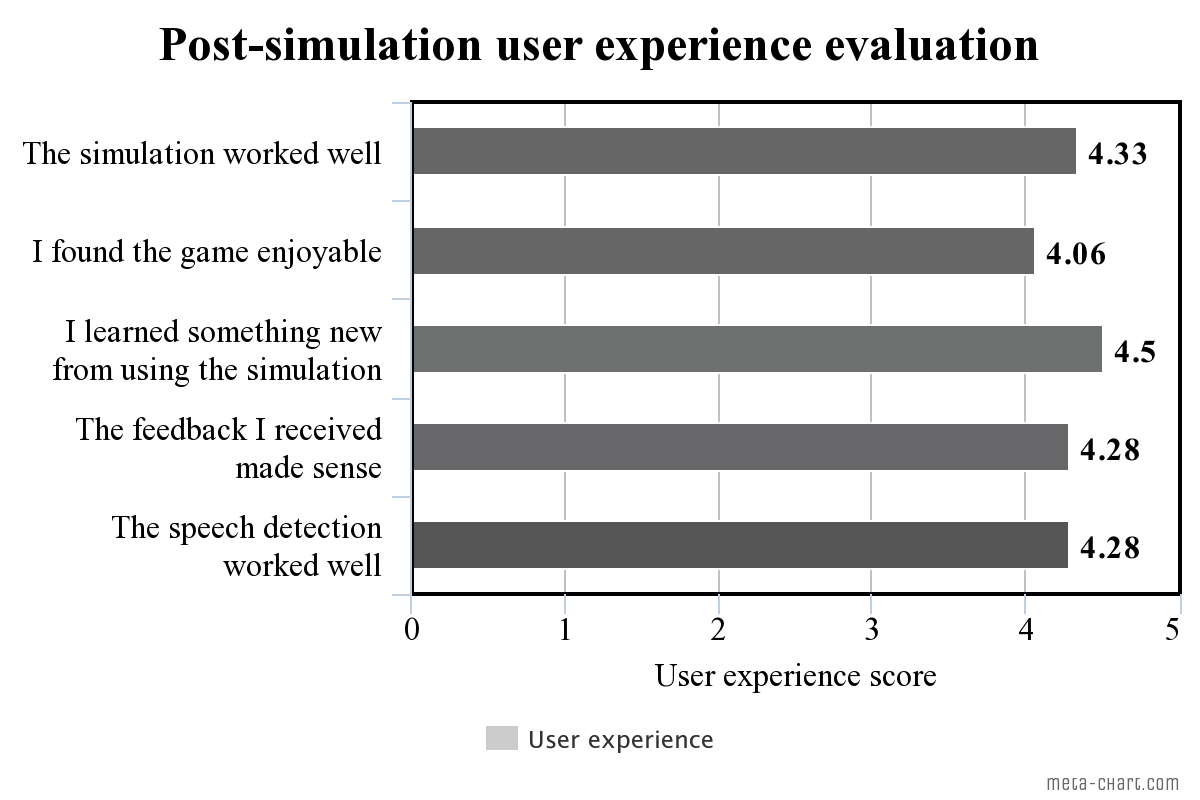}
    \caption{User experience survey result}
    \label{fig:survey}
\end{figure}

\textbf{\textit{H4:}} We used post simulation survey to evaluate the users' experience. We issued questionnaire to the participants after completing the scenarios in the simulation to evaluate the performance of the system as integrated. Five (5) questions were asked to measure the performance of each component:

\begin{figure}[ht]
    \centering
    \includegraphics[width=0.45\textwidth]{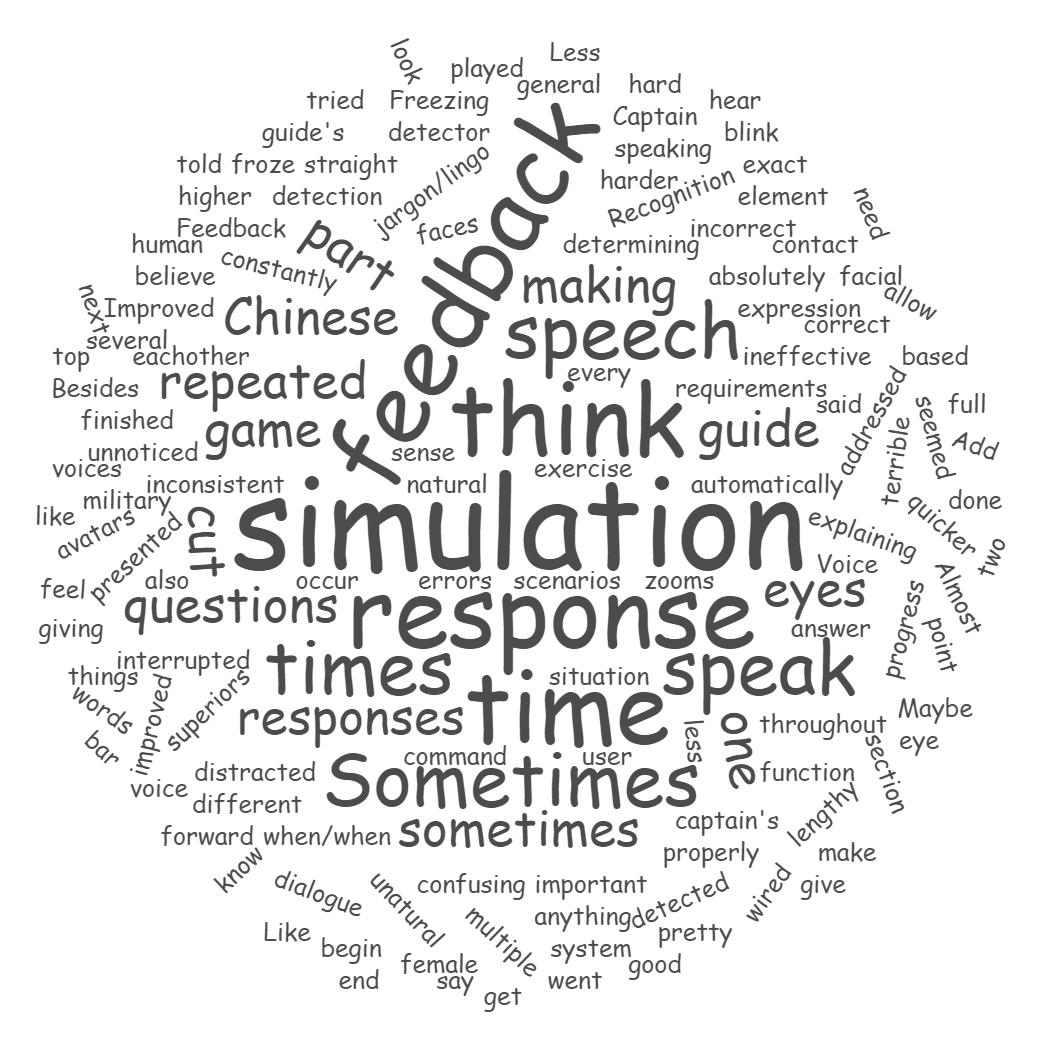}
    \caption{Word cloud of participants response }
    \label{fig:survey_words}
\end{figure}

figure \ref{fig:survey} showed that minimum of 80\% of the participants had positive response about each component. We also asked an open question where participants are to freely express what they think needs to be improved in the system. Responses to this question is as presented in the word cloud.

Some comments made by the participants in improving the simulation include; i) adding a progress bar; ii) improving the feedback system; and iii) Making the interaction more understandable to a non-military audience.

Appendix-I shows an example of a participant’s dialogue with the Chinese avatar during the simulation as well as the score received from the classification and the corresponding generated feedback.

\vspace{10pt}

\section{Conclusion}
In this paper we presented a data-driven simulation system for cultural awareness training. We improved previous simulation systems by implementing a spoken dialogue system in place of the menu-based option selection model. We implemented a data-driven multi-model cultural experts to evaluate and assign score to users' spoken responses based on abstracted cultural features. Our design also implemented an adaptive feedback mechanism that uses the score assigned by the expert module to serve the pedagogical role of informing the players how culturally appropriate their responses were. We evaluated the system by comparing the performance of the models with scores assigned by two human annotators. Our result showed that our data-driven intelligent experts models gave comparable performance with what human would adjudge similar utterance. The feature abstraction technique and adaptive feedback mechanisms also allowed us to solve the problem of manually scoring user's utterance. We also did not have to write feedback for every possible utterance that a user could make in the simulation. These components also allowed for real time adaptive feedback construction based on vector-score received by the learners' utterance. Most prior works have focused on developing techniques to improve the individual component of the intelligent tutoring system, our simulation focused on improving most of the modules to improve the realism of the interaction. While, providing explicit feedback in an interactive situation does not reflect the natural way in which learning occurs during interaction, our future works will explore techniques towards an adaptive response generation with implicit feedback. This will have the ultimate goal of improving the learning efficacy of cultural training simulation.

\section*{Acknowledgements}

The project is funded by the Army Research Lab under Research Funding Source Award Number: W911NF1820279. We thank the Data Science Institute of the University of Virginia for the support received for this project.

\bibliographystyle{IEEEtran} 
\bibliography{main}

\section{Appendix}
Below is the transcript of the dialogue of one of the participants for the first three scenes. We showed the score received (in parenthesis)\\ 
\textbf{Scene 1:}\newline
\textit{Captain Heist:} Good Morning. As you know, we are scheduled to meet our Chinese counterparts for the first time today. It is important that you make a good first impression. Unfortunately, I have had a requirement come up last minute and will not able to meet the Chinese captain today. I am having you go in my place. Please receive him cordially and begin making plans with him as soon as possible.\newline
\textit{Participant:} Thanks for the information.\newline
\textbf{Scene 2:}\newline
\textit{Participant:} Good morning Captain Wang.\newline
\textit{Captain Wang:} Good morning, Lieutenant.\newline
\textit{Participant:} Are you the leader of the Chinese component of the coalition.\newline
\textit{Captain Wang:} Yes, I am. I made arrangements with Captain Heist. Is he here?\newline
\textit{Participant:} Unfortunately Captain Heist had prior duties and I will be taking his place.\newline
\textit{Captain Wang:} Why couldn't he come?\newline
\textit{Participant:} He unfortunately had a prior engagements with other officers.\newline
\textit{Captain Wang:} I understand. Let’s begin preparing because we need to leave soon.\newline
\textit{Participant:} Great let's get started sir.\newline
\textbf{Scene 3:}\newline
\textit{Participant:} Captain Wang is there anything I can help you with?\newline
\textit{Captain Wang:} I'm Fine, Lieutenant. I just had some questions about the mission.\newline
\textit{Participant:} Well what are your concerns officer.\newline
\textit{Captain Wang:}	Well, my team and I were wondering what kind of supplies your team is planning on bringing.\newline
\textit{Participant:} Well Captain we haven't worked out all the Kinks but if you would like to meet are at our briefing later this afternoon you're more than welcome to come.\newline
\textit{Captain Wang:} I completely understand, but even some basic information would be useful for me to start my planning.\newline
\textit{Participant:} Well Captain we haven't worked out all the Kinks but if you would like to meet are at our briefing later this afternoon you're more than welcome to come.\newline
\textit{Captain Wang:} Please Lieutenant, it would help me a great deal if I knew exactly what supplies you have for the mission.\newline
\textit{Participant:} Well caps and we plan on bringing food water and other Hospital related equipment such as medicine to the site but we aren't sure what sort of Weaponry will be using but that will be discussed in the brief.\newline
\textit{Captain Wang:} Is there any way for me to get a look at the supply depot where you keep everything? I don't want to bother you any longer, so if it will save you time, I'll walk over myself.\newline
\textit{Participant:} Well Captain the Supply Depot is down the way to the left so if you want to check there that would be a good hint to so it will bring.\newline
\textit{Captain Wang:} Oh, well thank you Lieutenant! Good luck on the planning.\newline
\textit{Participant:}	I will be around until the mission brief this afternoon if you need me at any point.\newline

\end{document}